\documentclass[%
 reprint,
 superscriptaddress,
 amsmath,amssymb,
 aps,
 prl, 
 twocolumn
]{revtex4-1}

\usepackage{graphicx}
\usepackage{dcolumn}
\usepackage{epstopdf}
\usepackage{bm}

\usepackage[T1]{fontenc}
\usepackage{color}
\usepackage{amsthm}
\usepackage{amssymb}
\usepackage{esint}
\usepackage{comment}
\usepackage{hyperref}

\makeatletter

\newtheorem{thm}{\protect\theoremname}
  \theoremstyle{plain}
  \newtheorem{lem}[thm]{\protect\lemmaname}
  \theoremstyle{remark}
  
  \theoremstyle{plain}
  \newtheorem*{lem*}{\protect\lemmaname}
  \theoremstyle{plain}
  \newtheorem{prop}[thm]{\protect\propositionname}
  \theoremstyle{plain}

\usepackage[USenglish]{babel}
  \providecommand{\corollaryname}{Corollary}
  \providecommand{\lemmaname}{Lemma}
  \providecommand{\propositionname}{Proposition}
  \providecommand{\remarkname}{Remark}
\providecommand{\theoremname}{Theorem}

\begin{document}

\global\long\def\ve{\varepsilon}
\global\long\def\R{\mathbb{R}}
\global\long\def\Rn{{\mathbb{R}^{n}}}
\global\long\def\Rd{{\mathbb{R}^{d}}}
\global\long\def\E{\mathbb{E}}
\global\long\def\P{\mathbb{P}}
\global\long\def\bx{\mathbf{x}}
\global\long\def\vp{\varphi}
\global\long\def\ra{\rightarrow}
\global\long\def\smooth{C^{\infty}}
\global\long\def\symm{\mathcal{S}^n}
\global\long\def\psd{\mathcal{S}^n_{+}}
\global\long\def\pd{\mathcal{S}^n_{++}}
\global\long\def\dom{\mathrm{dom}\,}
\global\long\def\intdom{\mathrm{int}\,\mathrm{dom}\,}
\global\long\def\Tr{\mathrm{Tr}}

\newcommand{\ML}[1]{\textcolor{red}{[ML:#1]}}
\newcommand{\LL}[1]{\textcolor{blue}{[LL:#1]}}
\newcommand{\SuppStart}[1]{\textcolor{green}{[START possible supplemental]}}
\newcommand{\SuppEnd}[1]{\textcolor{green}{[END possible supplemental]}}

\newcommand{\bvec}[1]{\mathbf{#1}}
\renewcommand{\Re}{\mathrm{Re}}
\renewcommand{\Im}{\mathrm{Im}}
\newcommand{\textred}[1]{\textcolor{red}{#1}}

\newcommand{\mc}[1]{\mathcal{#1}}
\newcommand{\mf}[1]{\mathfrak{#1}}
\newcommand{\mcV}{\mathcal{V}}
\newcommand{\Vin}{V_{\mathrm{in}}}
\newcommand{\Vstar}{V^{\ast}}
\newcommand{\Jstar}{J_{\ast}}
\newcommand{\tJstar}{\wt{J}_{\ast}}
\newcommand{\Vout}{V_{\mathrm{out}}}
\newcommand{\RPA}{\mathrm{RPA}}
\newcommand{\xc}{\mathrm{xc}}
\renewcommand{\vr}{\bvec{r}}
\newcommand{\vF}{\bvec{F}}
\newcommand{\vg}{\bvec{g}}
\newcommand{\vR}{\bvec{R}}
\newcommand{\vq}{\bvec{q}}
\newcommand{\vx}{\bvec{x}}
\newcommand{\ud}{\,\mathrm{d}}
\newcommand{\ext}{\mathrm{ext}}
\newcommand{\KS}{\mathrm{KS}}
\newcommand{\Exc}{E_{\mathrm{xc}}}
\newcommand{\Vxc}{\hat{V}_{\mathrm{xc}}}
\newcommand{\Vion}{\hat{V}_{\mathrm{ion}}}
\newcommand{\abs}[1]{\lvert#1\rvert}
\newcommand{\norm}[1]{\lVert#1\rVert}
\newcommand{\average}[1]{\left\langle#1\right\rangle}
\newcommand{\wt}[1]{\widetilde{#1}}

\newcommand{\etc}{\textit{etc.}~}
\newcommand{\etal}{\textit{et al}~}  
\newcommand{\ie}{\textit{i.e.}~}
\newcommand{\eg}{\textit{e.g.}~}
\newcommand{\Or}{\mathcal{O}}
\newcommand{\mcF}{\mathcal{F}}
\newcommand{\lmin}{\lambda_{\min}}
\newcommand{\lmax}{\lambda_{\max}}
\newcommand{\Ran}{\text{Ran}}
\newcommand{\I}{\imath} 
\newcommand{\EE}{\mathbb{E}}
\newcommand{\NN}{\mathbb{N}}
\newcommand{\RR}{\mathbb{R}}
\newcommand{\CC}{\mathbb{C}}
\newcommand{\ZZ}{\mathbb{Z}}
\newcommand{\Hper}{H^1_\#(\Omega)}
\newcommand{\jmp}[1]{\jl#1\jr}
\newcommand{\al}{\{\hspace{-3.5pt}\{}
\newcommand{\ar}{\}\hspace{-3.5pt}\}}
\newcommand{\avg}[1]{\al#1\ar}
\newcommand{\jl}{[\![}
\newcommand{\jr}{]\!]}
\newcommand{\VN}{\mathbb V_N}

\title{Variational Structure of Luttinger-Ward Formalism and \\
Bold Diagrammatic Expansion for Euclidean Lattice Field Theory}

\author{Lin Lin}
\email{linlin@math.berkeley.edu}
\affiliation{Department of Mathematics, University of California,
Berkeley, California 94720, United States}
\affiliation{Computational Research Division, Lawrence Berkeley
National Laboratory, Berkeley, California 94720, United States}

\author{Michael Lindsey}
\email{lindsey@math.berkeley.edu}
\affiliation{Department of Mathematics, University of California,
Berkeley, California 94720, United States}

\begin{abstract}
The Luttinger-Ward functional was proposed more than five decades ago to
provide a link between static and dynamic quantities in a quantum many-body system. Despite its widespread usage,  the derivation of the
Luttinger-Ward functional remains valid only in the formal sense, and
even the very existence of this functional has been challenged by
recent numerical evidence.  In a simpler and yet highly relevant
regime, namely the Euclidean lattice field theory, we rigorously prove that
the Luttinger-Ward functional is a well defined universal functional
over all physical Green's functions.  Using the Luttinger-Ward
functional, the free energy can be variationally minimized with respect
to Green's functions in its domain. We then derive the widely used bold
diagrammatic expansion rigorously, without relying on formal arguments
such as partial resummation of bare diagrams to infinite order.
\end{abstract}

\pacs{Valid PACS appear here}
\maketitle

The Luttinger-Ward (LW) formalism~\cite{LuttingerWard1960} is an important component of the
Green's function theory of quantum many-body physics. The LW 
functional $\Phi[G]$ is a universal functional of the single-particle
Green's function $G$ and depends only on the form of the many-body
interaction, rather than any external field acting on a single particle.
The LW functional provides a link between static and dynamic
quantities,
in the sense that the Gibbs free energy $\Omega[G]$ can be
expressed as a functional of frequency-dependent $G$. 
The first derivative of the LW
functional gives the self-energy $\Sigma[G]=\frac{\delta \Phi}{\delta
G}[G]$.  The physical Green's function
satisfies the Dyson equation, which describes the stationary points of
the free energy. In the perturbative regime, the LW formalism 
provides the ``bold diagrammatic expansion''~\cite{FetterWalecka2003} for both $\Sigma[G]$ and
$\Phi[G]$ to all orders of diagrams, and thus
reduces the arbitrariness of the design of partially resummed perturbation
theories. 
In the non-perturbative regime, the LW formalism has been used to justify the
dynamical mean field theory
(DMFT)~\cite{GeorgesKotliarKrauthEtAl1996,Potthoff2006}.

Despite its success in both physics and
chemistry~\cite{DahlenVanVon2005,Ismail-Beigi2010,BenlagraKimPepin2011,RentropMedenJakobs2016},
the LW formalism remains justified only
in the formal sense.  This is a serious issue both in theory and in
practice. In fact, even the very existence of the LW functional is under
debate, with numerical evidence to the contrary appearing in the past
few
years for fermionic systems~\cite{KozikFerreroGeorges2015,Elder2014,TarantinoRomanielloBergerEtAl2017,GunnarssonRohringerSchaeferEtAl2017}.
One major difficulty lies in the fact that in quantum many-body physics
(fermionic or bosonic), $G$ is a frequency dependent quantity with
complicated singularity structures in the complex plane.  Even the
domain of the functional $\Phi[G]$ is unknown, and this forbids any
further rigorous discussion of its properties.

In this Letter, we present the first rigorous theory of the LW formalism
in the context of the Euclidean lattice field theory (such as the
$\varphi^{4}$ theory~\cite{AmitMartin-Mayor2005,Zinn-Justin2002}). In
contrast to the quantum many-body setting, Euclidean lattice field theories avoid a key theoretical challenge mentioned above in that the Green's function, defined as a two-point
correlator function, has a clear domain as the set of positive
semi-definite matrices with dimension equal to the size of the discretized system. Nonetheless, there is an exact
correspondence between the Feynman diagrammatic expansion of lattice
field theory and that of quantum many-body
physics~\cite{FetterWalecka2003,AmitMartin-Mayor2005}. Hence lattice field theory reveals
valuable information such as the \textit{structural properties} of diagrammatic expansions.  

For a general interaction form (not necessarily the quartic interaction),
we prove using Legendre duality that there exists a
universal functional of the Green's function, denoted
$\mathcal{F}[G]$, which is defined via a constrained minimization problem similar in spirit to that of the Levy-Lieb construction in
density functional theory~\cite{Levy1979,Lieb1983} at zero temperature
and the Mermin functional~\cite{Mermin1965} at finite temperature, and
further the ``density matrix functional theory'' developed
in~\cite{Baerends2001,SharmaDewhurstLathiotakisEtAl2008,BloechlPruschkePotthoff2013}.
We identify a natural one-to-one correspondence
between the interacting Green's function $G$ and the inverse
$G_{0}^{-1}$ of the non-interacting Green's function.
The LW functional $\Phi[G]$ is defined by subtracting a
logarithmically divergent component from $\mathcal{F}[G]$ . The functional derivative of the LW functional
defines the self-energy and is also universal. The free energy can
be expressed variationally as a minimum over all physical Green's functions, and the
self-consistent solution of the Dyson equation yields its global and
unique minimizer.  Finally, using the LW formalism for quartic
interactions, we rigorously recover the form of the bold diagrammatic
expansion appearing in the quantum many-body setting, without any
reference to the non-interacting Green's function $G_{0}$.
This Letter gives our main results and outlines of the proofs, and we
refer readers to an upcoming full length publication for the rigorous presentation
of the results~\cite{MBPTMath}.

After proper discretization, a Euclidean lattice field theory can be described by the partition function
\begin{equation}
Z=\int_{\Rn}e^{-\frac{1}{2}x^{T}Ax-U(x)}\ud x.
\label{eqn:Zgeneral}
\end{equation}
For instance, the partition function of a scalar $\varphi^{4}$ theory in
a $d$-dimensional space is 
\begin{equation}
  Z = \int \mathcal{D} \varphi(\vr) 
  e^{-\int_{\Rd} \frac12 \abs{\nabla\varphi(\vr)}^2 + a \varphi^2(\vr) + u
  \varphi^{4}(\vr) \ud \vr}.
  \label{eqn:Zphi4}
\end{equation}
After discretizing the field $\varphi(\vr)$ on a lattice of size
$n$ with components denoted by $\{x_{i}\}_{i=1}^{n}$, we can
rewrite the quadratic part in
Eq.~\eqref{eqn:Zphi4} by a quadratic form given by a symmetric matrix $A$ and the quartic part by a
polynomial $U(x)$ as in Eq.~\eqref{eqn:Zgeneral}. 
Here $A$ can be associated
with the non-interacting Hamiltonian in quantum many-body physics, and
$U$ with the interaction term. 
The form in Eq.~\eqref{eqn:Zgeneral} is very general and can represent
interaction terms that are quartic, beyond quartic, or even
non-polynomial in classical and quantum statistical mechanics (see
e.g.~\cite{FeynmanHibbs1965,AmitMartin-Mayor2005,Wilson1975}). In order for
the integral to be well defined, we assume that $U(x)$ goes to infinity
faster than any quadratic function of $x$. 

Let $\mathcal{S}^{n}$, $\mathcal{S}^{n}_+$, and $\mathcal{S}^{n}_{++}$
denote respectively the sets of symmetric, symmetric positive
semidefinite, and symmetric positive definite $n\times n$ matrices, so
$\mathcal{S}^{n}_{++}\subset\mathcal{S}^{n}_+\subset\mathcal{S}^{n}$. Then the partition function $Z$ in Eq.~\eqref{eqn:Zgeneral} can be viewed as
a functional of $A\in \mathcal{S}^{n}$ denoted by $Z[A]$. The Gibbs free
energy is defined as $\Omega[A]:=-\log Z[A]$, and $\Omega[A]$ is
strictly concave and $\smooth$-smooth on $\symm$.

The derivative of $\Omega$ with respect to $A_{ij}$ defines the
two-point correlator, or the Green's
function
\begin{equation}
  G_{ij}:=\nabla_{ij}\Omega[A]=\frac{1}{Z[A]}\int x_i x_j \,e^{-\frac{1}{2}x^{T}Ax-U(x)}\ud x.
  \label{eqn:Greendef}
\end{equation}
The matrix $G\in \symm$ is the two-point correlator with respect to the
probability distribution
$\rho(x)=e^{-\frac{1}{2}x^{T}Ax-U(x)}/Z[A]$, and hence
$G\in\mathcal{S}^{n}_{++}$. $G$ is called the interacting Green's
function, and if we set $U(x) \equiv 0$ we obtain the non-interacting Green's
function $G_0 = A^{-1}$. However, the non-interacting Green's function
$G_{0}$ is only well defined for $A\in \mathcal{S}^{n}_{++}$, while the
interacting Green's function $G$ is well defined for any $A\in
\mathcal{S}^{n}$ owing to the growth property of the interaction term
$U(x)$. 

Let $\mathcal{M}$ be
the space of probability density functions on $\Rn$ with moments up to second order. 
Define $\mathcal{G}:\mathcal{M} \ra \mathcal{S}_{++}^{n}$ by
$\mathcal{G}(\rho)=\int
xx^T\,\rho\ud x$, which maps a probability density to a
Green's function in $\mathcal{S}_{++}^{n}$. On the other hand, for any
$G\in\mathcal{S}_{++}^{n}$, it is clear that the inverse mapping
$\mathcal{G}^{-1}(G)$ is a non-empty set through the construction of a Gaussian
distribution. For general interaction $U(x)$,
our main results are given in Theorem~\ref{thm:variation}
and~\ref{thm:lwfunctional}.

\begin{thm}[Variational structure]\label{thm:variation}
  The Gibbs free energy can be expressed variationally via
  \begin{equation}
    \Omega[A]=\inf_{G\in\mathcal{S}^{n}_{++}}\left(\frac{1}{2}\mathrm{Tr}[AG]-\mathcal{F}[G]\right),\label{eq:omegaInf2}
  \end{equation}
  where 
  \begin{equation}
\mathcal{F}[G]:=\sup_{\rho \in \mathcal{G}^{-1}(G)}\left[S(\rho)-\int U\,\rho\ud x\right] \label{eq:Fdef}
\end{equation}
  is the Legendre dual of $\Omega[A]$ with respect to the inner product
  $\langle A,G\rangle = \frac{1}{2}\mathrm{Tr}[AG]$.  Here 
  $S:\mathcal{M} \ra \R$ is the entropy and is defined as $S(\rho)=
  -\int \rho \log \rho\ud x$.  The mapping $G[A]:=\nabla\Omega[A]$ is a
  \textit{bijection} $\symm \ra \pd$, with inverse given by
  $A[G]:=\nabla\mathcal{F}[G]$.
\end{thm}

Note that $\mc{F}$ depends only on $G$ and, implicitly, on the interaction $U$, but it is independent of the
non-interacting term $A$. In this sense, $\mc{F}$ is a \textit{universal} functional of the Green's
function $G$.

The last statement of Theorem~\ref{thm:variation} suggests that
$\nabla\mathcal{F}$ (hence also $\mathcal{F}$) should approach infinity as $G$ approaches the
boundary of $\mathcal{S}_{++}^{n}$.  Remarkably, we can explicitly
separate the part that accounts for the blowup of $\mathcal{F}$ at the
boundary. 
Consider the case in which $U\equiv0$, and $\mathcal{F}[G]=\sup_{\rho\in
\mathcal{G}^{-1}(G)}S(\rho)$. The maximizer is attained by the probability density
corresponding to the Gaussian distribution $\mathcal{N}(0,G)$. Hence
\begin{equation}
  \mathcal{F}[G]=\frac{1}{2}\log\left((2\pi e)^{n}\det G\right)=\frac{1}{2}\mathrm{Tr}[\log(G)]+\frac{n}{2}\log(2\pi e),
  \label{}
\end{equation}
which diverges logarithmically as $G$ approaches the
boundary of $\mathcal{S}^{n}_{++}$. Subtracting away this singular part,  we define
the \textit{Luttinger-Ward (LW) functional} as
\begin{equation}
  \Phi[G]:=2\mathcal{\mathcal{F}}[G]-\mathrm{Tr}[\log(G)]-\Phi_{0},
  \quad \Phi_{0}=n\log(2\pi e).
  \label{eq:LWdef}
\end{equation}

\begin{thm}[Luttinger-Ward functional]\label{thm:lwfunctional}
  The LW functional in Eq.~\eqref{eq:LWdef} is universal, satisfies $\Phi[G]\equiv0$
  for non-interacting systems, and extends
  continuously up to the boundary of $\mathcal{S}_{++}^{n}$. The self-energy functional is defined as $\Sigma[G]=\nabla \Phi[G]$ and is also
  universal. The solution of the Dyson equation
  \begin{equation}
    G^{-1}=A-\Sigma[G]
    \label{eq:dyson}
  \end{equation}
  in $S^{n}_{++}$ is the unique minimizer of the free energy in~\eqref{eq:omegaInf2}.
\end{thm}

According to the preceding discussion, for $A\in
\mathcal{S}^{n}_{++}$, we have $G_{0}=A^{-1}$, and the Dyson
equation~\eqref{eq:dyson} can be written equivalently as
\begin{equation}
  G=G_{0}+G_{0}\Sigma[G]G.
  \label{eq:dyson2}
\end{equation}
This is the common starting point for deriving the Feynman
diagram expansion~\cite{FetterWalecka2003,AmitMartin-Mayor2005} with propagator
$G_{0}$, i.e. the ``thin-line'' (or ``bare'') diagrammatic expansion. In our setting,
this expansion is meaningless when $A\notin \mathcal{S}^{n}_{++}$, since
the corresponding partition
function~\eqref{eqn:Zgeneral} diverges in the non-interacting limit. On
the other hand, the Dyson equation in the form of~\eqref{eq:dyson} is
more general and is valid for any $A\in \mathcal{S}^{n}$.

When the self-energy functional $\Sigma[G]$ is known,
Eq.~\eqref{eq:dyson} can be solved to obtain $G$. On the other hand,
Eq.~\eqref{eq:dyson} can also be used in the reverse direction to
compute $\Sigma$ once $A$ and an approximation to $G$ are available.
This is the approach taken in DMFT~\cite{GeorgesKotliarKrauthEtAl1996},
which approximates $\Sigma$ by solving a number of impurity problems on
local domains. 
This second use of the Dyson equation seems to suggest that $\Sigma$
depends on both $G$ and $A$, though we have claimed it to be a universal
functional of $G$! However, the one-to-one mapping between $A$ and $G$
furnished by Theorem~\ref{thm:variation} resolves this paradox, and
$\Sigma[G]$ is indeed well defined for the Euclidean lattice field theory.  A
similar correspondence for many-body quantum systems is still under
debate~\cite{Potthoff2006,KozikFerreroGeorges2015,Elder2014,TarantinoRomanielloBergerEtAl2017,GunnarssonRohringerSchaeferEtAl2017}.

Though the dependence of the LW functional on the interaction
$U$ was only implicit in the preceding discussion, we may explicit consider
this dependence, including it in our notation as $\Phi[G;U]$. The
same convention will be followed for other functionals without comment. 
As we shall see, unlike the functional $\mathcal{F}[G]$, which diverges
at the boundary of $G$,  the LW functional $\Phi[G;U]$ extends
continuously to the boundary of $G$. This relates to the possibility of establishing a diagrammatic expansion $\Phi[G;U]$ with respect to
the interaction strength. 

So far we have considered the LW formalism for any interaction that
satisfies a sufficient growth condition.  In order to draw a
closer connection with the diagrammatic expansion used in quantum
many-body physics, we now restrict our attention to the quartic interaction
\begin{equation}
  U(x) = \frac{1}{8} \sum_{i,j=1}^{n} v_{ij} x_{i}^2 x_{j}^2.
  \label{eqn:Uterm}
\end{equation}
Here $v_{ij}=v_{ji}$ is symmetric, and the symmetry factor $8$
simplifies the counting when deriving diagrammatic approximations.
The interaction~\eqref{eqn:Uterm} can mimic a short range interaction
as well as a long range (such as Coulomb) interaction in its second
quantized form~\cite{FetterWalecka2003}. One can derive an exact correspondence between the
Feynman diagrammatic expansions in this lattice field theory and
those in condensed matter physics~\cite{AmitMartin-Mayor2005},
neglecting the particle-hole distinction.

For fixed interaction $U$ and $G\in
\mathcal{S}_{++}^n$, we may define a perturbative expansion of
$\Phi[G;\ve U]$ with respect to the interaction strength $\ve$.
Theorem~\ref{thm:bolddiagram} shows that the bold diagrammatic expansion
of the LW functional at $G$ can be understood as the asymptotic series in
$\ve$ for the LW functional at $G$. The importance of this theorem is
to show that the bold diagrammatic expansion is well defined without any
reference to the non-interacting Green's function $G_{0}$.

\begin{thm}[Bold diagrammatic expansion]\label{thm:bolddiagram}
 For a general interaction $U$, the
  Luttinger-Ward functional and the self-energy have the following asymptotic series expansions
  \begin{equation}
    \Phi [G;\ve U] = \sum_{k=1}^\infty \Phi^{(k)}[G;U]\, \ve^k, 
    \Sigma [G;\ve U] = \sum_{k=1}^\infty \Sigma^{(k)}[G;U]\, \ve^k.
    \label{eqn:asymptot}
  \end{equation}
 Moreover, for $U$ of the form~\eqref{eqn:Uterm}, the coefficients of the asymptotic series satisfy
  \begin{equation}
    \Phi^{(k)}[G;U] = \frac{1}{2k} \Tr\left[G \Sigma^{(k)}[G;U]\right],
    \label{eqn:PhiSigmaBold}
  \end{equation}
  and $\Sigma^{(k)}[G;U]$ consists of all one-particle irreducible skeleton diagrams of order $k$. 
\end{thm}

\begin{figure}[h]
  \begin{center}
    \includegraphics[width=0.45\textwidth]{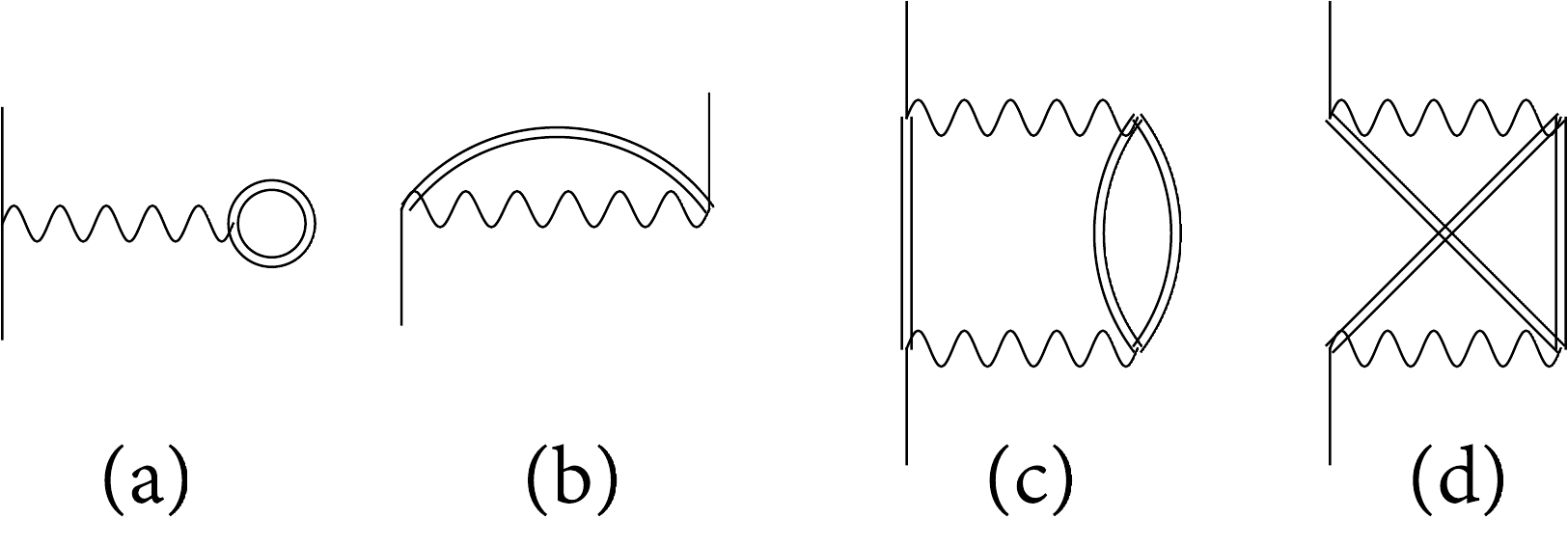}
  \end{center}
  \vspace{-2em}
  \caption{Bold diagrams for the first order (a-b) and second order
  (c-d) contribution to the self energy for interaction of the
  form~\eqref{eqn:Uterm}.}
  \label{fig:feynmanSigmaG2}
\end{figure}

Note that for the series in Eq. \eqref{eqn:asymptot} to be asymptotic
means that the error of the $N$-th partial sum is $O(\ve^{N+1})$ as $\ve
\ra 0$. For concreteness, Fig.~\ref{fig:feynmanSigmaG2} gives the bold
diagrammatic
expansion for the self energy up to the second order, where
\begin{equation*}
  \small
  \begin{split}
  \left(\Sigma^{(1)}[G]\right)_{ij} = &
  -\frac12 \left(\sum_{k} v_{ik} G_{kk}\right)
  \delta_{ij} - v_{ij} G_{ij}, \\
  \left(\Sigma^{(2)}[G]\right)_{ij} = &
  \frac12 G_{ij} \left(\sum_{k,l} v_{ik} G^2_{kl}
  v_{lj}\right) + \sum_{k,l} v_{ik} G_{kj} G_{kl} G_{li} v_{jl},
  \end{split}
  \label{}
\end{equation*}
and the $U$-dependence is determined by $v_{ij}$ as in
Eq.~\eqref{eqn:Uterm}.
Compared to the Feynman diagrams for condensed matter systems, we find
that \textit{not} coincidentally, Fig.~\ref{fig:feynmanSigmaG2} (a),(b) correspond to the Hartree and
Fock exchange diagrams, respectively, and (c),(d) correspond to the ring and second-order exchange diagrams, respectively. The only difference is that the
lines in the diagrams of Fig.~\ref{fig:feynmanSigmaG2} do not possess
directions, due to the absence of any distinction between creation and
annihilation operators. 

Interestingly, the relation~\eqref{eqn:PhiSigmaBold} was
originally assumed to be true to obtain a formal derivation of the LW
functional~\cite{LuttingerWard1960,MartinReiningCeperley2016}.  Our
proof here does not rely on such formal manipulation, but instead only on
the transformation rule (Proposition~\ref{prop:LWtransformation}) below and the quartic nature of the
interaction $U$.

Finally we remark that certain properties in the Euclidean setting, such
as the concavity of the free energy functional, can noticeably fail in
the non-Euclidean setting. Indeed, the original setting for the
LW formalism is a field theory described by the fermionic coherent
state path integral represented by Grassmann variables. The free energy
functional is non-concave and the induced Legendre correspondence may not
be one-to-one. This leads to the failure of the LW formalism observed in
~\cite{Potthoff2006,KozikFerreroGeorges2015,Elder2014,TarantinoRomanielloBergerEtAl2017,GunnarssonRohringerSchaeferEtAl2017}, and the full picture of the LW functional remains to be revealed.  
Intriguingly, the LW formalism may also be seen as an expansion of a
static density matrix
formalism~\cite{Mermin1965,BloechlPruschkePotthoff2013}, which itself \emph{does} enjoy convexity
properties, and hence well-defined Legendre duality.
However, 
the density matrix formalism is \emph{not} induced in the same way by a field theory
and does not enjoy even formally properties  such as the diagrammatic expansion.
The Euclidean field theory setting can then be viewed as combining the
best of both worlds, in that
it enjoys both the convexity properties needed for the non-perturbative
definition of the Legendre dual functionals, as well as the formal
properties convenient for systematic approximation as in diagrammatic
expansions and DMFT.

This work also opens up several immediate research directions. By
making a connection between quantum many-body physics and Euclidean lattice field
theory, it lowers the barrier for quantitatively assessing the 
effectiveness of bold diagrammatic schemes and other numerical schemes based on many-body
perturbation theory such as the $GW$ theory~\cite{Hedin1965}.
Possible topics to be developed include the effectiveness of self-consistent many-body
perturbation theories and the effectiveness of the vertex correction
methods in the $GW\Gamma$ theory. Theoretical properties of impurity models and embedding
schemes, such as the dynamical mean field theory (DMFT), can also be
studied in the context of Euclidean lattice field theory using the LW
formalism. 

\vspace{1em}
\vspace{1em}
\textit{Outline of the proof of Theorem \ref{thm:variation}:} 
First, we reformulate the computation of the Gibbs free energy
$\Omega[A]=-\log Z[A]$ as a minimization problem: 
\begin{equation}
\Omega[A]=\inf_{\rho\in\mathcal{M}}\left[\int\left(\frac{1}{2}x^{T}Ax+U(x)\right)\,\rho\ud x -S(\rho)\right].\label{eq:omegaInf1}
\end{equation}
This is the classical Gibbs variational principle.
Next, we split the infimum
in Eq.~\eqref{eq:omegaInf1} as
\begin{equation}
\Omega[A]=\inf_{G\in\mathcal{S}_{++}^{n}}\left(\frac{1}{2}\mathrm{Tr}[AG]+\inf_{\rho\in \mathcal{G}^{-1}(G)}\left[\int U\,\rho\ud x-S(\rho)\right]\right),
  \label{}
\end{equation}
Here we have used $\int x^{T}Ax\,\rho \ud x =\mathrm{Tr}[\mathcal{G}(\rho)A]$. By
introducing the functional~\eqref{eq:Fdef}, we obtain the variational
formulation in~\eqref{eq:omegaInf2}.
Now Eq.~\eqref{eq:omegaInf2} means precisely that
$\Omega$ is the Legendre dual, or more precisely the \emph{concave
conjugate} of $\mathcal{F}$ with respect to the inner product $\langle
A,G\rangle = \frac{1}{2}\mathrm{Tr}[AG]$~\cite{SuppMat}.
This is denoted by $\Omega=\mathcal{F}^{*}$.

One can further prove that $\mathcal{F}$ is concave on $\pd$ and diverges to $-\infty$ at the boundary of $\pd$ ~\cite{SuppMat,MBPTMath}. 
Based on these facts, we have that
$\mathcal{F}=\mathcal{F}^{**}$, i.e., $\mathcal{F}= \Omega^{*}$, so $\mathcal{F}$ and
$\Omega$ are concave duals of one another. Furthermore, it can be shown using the strict concavity and $\smooth$-smoothness of $\Omega = \mathcal{F}^*$ that
$\mathcal{F}$ is strictly concave and $\smooth$-smooth on
$\pd$.

The Legendre duality suggests that $\nabla \mathcal{F}$ and $\nabla \Omega$ are inverses of one another, i.e., the mapping $G[A]:=\nabla\Omega[A]$
is a \textit{bijection} $\symm \ra \pd$, with inverse
given by $A[G]:=\nabla\mathcal{F}[G]$. Moreover, we remark that 
for any $G\in \mathcal{S}_{++}^{n}$, the supremum in the definition \eqref{eq:Fdef} of $\mathcal{F}[G]$ is
uniquely attained at define the probability density 
$\rho_{G}(x) := \frac{1}{Z[A[G]]} e^{-\frac{1}{2}x^{T}A[G]x-U(x)}$.
Q.E.D.

\vspace{1em}
\textit{Outline of the proof of Theorem~\ref{thm:lwfunctional}:} 
The differentiability
of the LW functional on $\pd$ directly follows from the
$\smooth$-smooth property of $\mathcal{F}[G]$ on $\pd$.
Hence the self-energy $\Sigma[G]=\nabla \Phi[G]$ is well-defined on $\pd$. Using
the LW functional, Eq.~\eqref{eq:omegaInf2} can be written
as
\begin{equation}
    \Omega[A]=\frac12
    \inf_{G\in\mathcal{S}^{n}}\left(\mathrm{Tr}[AG]-\mathrm{Tr}[\log(G)]-\Phi[G]-\Phi_{0}\right),
  \label{eq:omegaInf3}
\end{equation}
The Euler-Lagrange equation with respect to $G$ gives the Dyson
equation~\eqref{eq:dyson}, and the uniqueness of the solution follows
from that of the minimizer in Theorem~\ref{thm:variation}.

We now establish that unlike
$\mathcal{F}[G]$, which blows up at the boundary of $\pd$, the LW
functional $\Phi[G]$ extends continuously to the boundary of
$\mathcal{S}^{n}_{++}$, so in fact the LW functional is well-defined on $\psd$. 
We first state a useful property of the LW
functional, which relates a basis transformation of the Green's
function with a transformation of the interaction~\cite{SuppMat}.

\begin{prop}[Transformation rule]
\label{prop:LWtransformation}Let $G\in\mathcal{S}_{++}^{n}$, and
let $U$ be the interaction term. Let $T$ denote an invertible matrix in
$\R^{n\times n}$, as well as the corresponding linear transformation
$\Rn\ra\Rn$. Then 
\[
\Phi[TGT^{*};U]=\Phi[G;U\circ T].
\]
 \end{prop}

Using the transformation rule, we only need to specify a formula for computing $\Phi[G]$ for matrices of the form
$G=\left(\begin{array}{cc}
  G_{p} & 0\\
0 & 0
\end{array}\right)$ 
with $G_{p}\in \mathcal{S}^{p}_{++}$, $p \leq n$.
Indeed, any matrix $G \in \psd$ with rank $p \leq n$ can be represented
as such after an appropriate change of basis, so together with the
transformation rule, such a formula will pin down the value of $\Phi[G]$
for all $G \in \psd$.

Define a $p$-dimensional interaction $U_p: \R^p \ra \R$ by the rule
$U_{p}(\,\cdot\,) = U(\,\cdot\,,0)$. We prove that~\cite{MBPTMath}
\begin{equation}
  \Phi_{n}\left[G;U\right]=\Phi_{p}\left[G_{p};U_{p}\right].
  \label{eqn:Phiboundary}
\end{equation}
Here $\Phi_n$ and $\Phi_p$ are the LW functionals for the $n$-dimensional and $p$-dimensional lattice field theories, respectively. Q.E.D.

\vspace{1em}
\textit{Outline of the proof of Theorem~\ref{thm:bolddiagram}:}
For a given Green's function $G$ and interaction $U$, for notational simplicity we will omit
the dependence on $G$ and $U$ from the notation via the definitions $\Phi(\ve)
:= \Phi [G;\ve U]$ and $\Sigma(\ve) = \Sigma[G;\ve U]$.
We first prove that~\footnote{This existence proof is non-constructive, and hence
the series coefficients still need to be determined.
See~\cite{MBPTMath}.} there \textit{exists} an asymptotic series
of the form~\eqref{eqn:asymptot}.
We abbreviate the notation
for the series coefficients via $\Phi^{(k)} := \Phi^{(k)}[G;U]$ and
$\Sigma^{(k)} := \Sigma^{(k)}[G;U]$.  Here the superscript $(k)$ is
just a notation and does not indicate the $k$-th order derivative.

Theorem \ref{thm:bolddiagram} then consists of identifying that these
coefficients are given by the bold diagrammatic expansion using $G$ and
$U$.  Our strategy is to first evaluate the expansion for the self-energy and then pin down the coefficients for the LW functional by
proving the relation \eqref{eqn:PhiSigmaBold}.
Since the series expansion is only valid in the asymptotic sense, for
any finite $\varepsilon$ we consider the truncation at finite order
$N$, which is denoted by $\overline{\Sigma}^{(N)}(\ve) := \sum_{k=0}^N
\Sigma^{(k)}\, \ve^k$. Then we have $\Sigma (\ve) -
\overline{\Sigma}^{(N)}(\ve) = O(\ve^{N+1})$.
Lemma~\ref{lem:SEpartialSum} says that $\overline{\Sigma}^{(N)}(\ve)$
can be identified as the \textit{exact} self energy with respect to a
modified interaction term~\cite{SuppMat}: 
\begin{lem}
\label{lem:SEpartialSum}
$\overline{\Sigma}^{(N)}(\ve)$ is the self-energy at $G$ induced by the
modified interaction ${U}^{(N)}_\ve(x) := \ve U(x) + \frac{1}{2} x^T
\left(\Sigma(\ve) - \overline{\Sigma}^{(N)}(\ve) \right) x $. In other
words, $\overline{\Sigma}^{(N)} (\ve) = \Sigma[G; {U}^{(N)}_\ve]$
is the exact self-energy corresponding to a non-interacting Green's function 
\begin{equation}
\label{eqn:nonintEps}
{G}_0^{(N)}(\ve) := 
\left( G^{-1} + \overline{\Sigma}^{(N)}(\ve) \right)^{-1},
\end{equation}
and the interaction $U_\ve ^{(N)}$.
\end{lem}

A difficulty in proving Theorem \ref{thm:bolddiagram} is that, although
we wish to employ the technique that resums bare self-energy diagrams to
bold self-energy diagrams, $\Sigma [G; \ve U]$ is defined without
reference to any bare propagator. Lemma \ref{lem:SEpartialSum} consists
of identifying a bare propagator $G_0^{(N)}$ that can generate (up to
negligible error) the bold propagator $G$ under the interaction $\ve U$.

Indeed, note that for any $x$, we have that $ U_\ve ^{(N)}(x) = \ve U(x) + O(\ve^{N+1})$. For the purpose of this analysis, $O(\ve^{N+1})$ error should be thought of as negligibly small. It can then be shown that $\overline{\Sigma}^{(N)}(\ve) = \widetilde{\Sigma}^{(N)}(\ve) + O(\ve^{N+1})$, where $\widetilde{\Sigma}^{(N)}(\ve)$ is the exact self-energy corresponding to non-interacting Green's function ${G}_0^{(N)}(\ve)$ and interaction $\ve U$, i.e., we can replace $U_\ve ^{(N)}$ by $\ve U$ with negligible effect on the self-energy.

By summing Green's function diagrams (with bare propagator ${G}_0^{(N)}(\ve)$ and interaction $\ve U$)  up to the \emph{finite} order $N$, we define a resummed bold propagator that differs from $G$ only by an error of $O(\ve^{N+1})$.  Then we may employ the standard combinatorial argument that the bold diagram expansion for the self-energy up to order $N$ accounts for all bare diagrams up to order $N$. It follows that $\widetilde{\Sigma}^{(N)}(\ve)$---and hence also $\overline{\Sigma}^{(N)}(\ve)$---is, up to an error of $O(\ve^{N+1})$, given by the bold diagram expansion up to order $N$ with bold line $G$ and interaction $\ve U$. But since this expansion and $\overline{\Sigma}^{(N)}(\ve)$ are both polynomials of order $N$ in $\ve$, it follows that $\overline{\Sigma}^{(N)}(\ve)$ is exactly given by the bold diagram expansion up to order $N$, as was to be shown. 
All that remains for the proof of Theorem \ref{thm:bolddiagram} is to
establish the relation \eqref{eqn:PhiSigmaBold}. This formula is a
consequence of the transformation rule~\cite{SuppMat}. Q.E.D.

This work was partially supported by the National Science Foundation
under Grant No. DMS-1652330, and by the Department of Energy under Grant
No. DE-SC0017867, No. DE-AC02-05CH11231 (L. L.), and by the National
Science Foundation Graduate Research Fellowship Program under grant
DGE-1106400 (M. L.).  We thank Fabien Bruneval, Roberto Car, Garnet
Chan, Lek-Heng Lim, Sohrab Ismail-Beigi, Nicolai Reshetikhin, and Chao
Yang for helpful discussions.

\bibliography{lwref}
\end{document}